# Lightweight Geometric Deep Learning for Molecular Modelling in Catalyst Discovery

Patrick Geitner
University of Rochester
pgeitner@ur.rochester.edu

## Abstract

*New technology for energy storage is necessary for the large-scale adoption of renewable energy sources like wind and solar. The ability to discover suitable catalysts is crucial for making energy storage more cost-effective and scalable. The Open Catalyst Project aims to apply advances in graph neural networks (GNNs) to accelerate progress in catalyst discovery, replacing Density Functional Theory-based (DFT) approaches that are computationally burdensome. Current approaches involve scaling GNNs to over 1 billion parameters, pushing the problem out of reach for a vast majority of machine learning practitioner around the world. This study aims to evaluate the performance and insights gained from using more lightweight approaches for this task that are more approachable for smaller teams to encourage participation from individuals from diverse backgrounds. By implementing robust design patterns like geometric and symmetric message passing, we were able to train a GNN model that reached a MAE of 0.0748 in predicting the per-atom forces of adsorbate-surface interactions, rivaling established model architectures like SchNet and DimeNet++ while using only a fraction of trainable parameters.*

## 1. Introduction

As the world transitions to reliance on renewable energy sources to combat the effects of climate change, new technology must be developed to allow for adoption on a large scale. Energy sources like wind and solar power are intermittent, so efficient methods for storing this energy is crucial for ensuring a stable and consistent supply of power. Current methods for energy storage typically involve converting renewable energy to other fuels, like hydrogen, but these chemical reactions are costly and require a large amount of energy [1]. Catalysts, which can lower the energy required for these chemical reactions, could make energy storage more efficient and cost-effective. Current simulations to identify suitable catalysts use density functional theory (DFT) to calculate the electronic structure of potential catalysts to predict their behavior in chemical reactions [2]. However, simulations using DFT are extremely computationally expensive, slowing the current rate of progress in this field.

### 1.1. The Open Catalyst Project

In recent years, geometric deep learning has made traction in other scientific applications involving molecular modeling with the generalization of deep neural networks to non-Euclidean domains such as graph representations [3].

To apply advances in graph neural networks (GNNs) to accelerate progress in catalyst discovery, Facebook AI Research (FAIR) and Carnegie Mellon University (CMU) created the Open Catalyst Project. This effort yielded a public dataset containing 1.3 million DFT relaxations consisting of over 260 million single-point evaluations to foster attempts to train deep learning models that can approximate DFT calculations with far less time and computational resources [4].

The resulting dataset contains information about the behavior of atoms and molecules in an adsorbate-surface system. More specifically, each observation provides the atomic structure of the adsorbate-surface system, the energy of the combined adsorbate-surface system, as well as the per-atom forces during the interaction. The task is to use the structure of the overall atomic system, represented as a graph, along with features describing the atoms, catalysts, and adsorbates, to predict the energy and per-atom forces in the system [1]. Learning this relationship would allow a model to extract important information about the properties of a surface and how it interacts with different molecules, including the electronic structure, which is important for understanding its catalytic activity. A successful model would allow researchers to identify suitable catalysts for a given reaction much more efficiently than using traditional DFT calculations. Figure 1 shows an example adsorbate-surface system from the dataset.

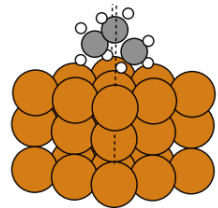

Figure 1: Propane (adsorbate) on a copper (catalyst) surface. By learning how the copper surface influences different molecules during adsorption, we can better understand copper's catalytic activity.

## 1.2. This Study

Given the complexity of this problem, much of the current literature has focused on very large models, with recent work from FAIR to scale GNNs to over one-billion parameters and over 11,000 GPU training hours for this task [4, 5]. Because of this, progress on this problem has been dominated by large AI research groups like FAIR and DeepMind, with very little engagement from individual machine learning practitioners and university laboratories. However, since progress in this domain is crucial for our ability to adopt renewable energy sources at scale, it is important that many people coming from a diverse set of backgrounds begin thinking about potential solutions.

In this study, we will evaluate the performance and insights gained from using more lightweight approaches for this task that are more approachable for individual machine learning practitioners around the world. To do this, we will use the smallest provided training split of the data, containing 200K observations (20GB storage) to train models with less than 5 million parameters for this task. To do this, we will use successful domain-specific model design choices but scale the models down to a fraction of the size to see how the predictive performance changes. This will also allow us to assess the utility of lightweight methods for determining design choices to devote computational resources to scaling up in the future.

## 2. Related Work

Some of the earliest traction in modeling the electronic structure of atoms, molecules, and solids was with the development of Density Functional Theory (DFT) with improvements in the 1990s that would be awarded the Nobel Prize in 1998 [2]. Calculations using DFT are used to extract information about a surface from which we can understand its catalytic activity. Unfortunately, single relaxation computations using DFT can take days, making it intractable to run experiments with many catalysts [6]. The Open Catalyst Project uses DFT calculations to generate the dataset used in this study in the hopes that deep learning architectures can be developed to replace DFT with far less computational burden.

Recent developments in the deep learning literature point to the success of GNNs in applications where data in not well-represented in Euclidean space [7]. This has led to several applications of GNNs in molecular modeling including predicting molecular properties, dynamics simulations, and de novo generation [8]. Several baseline model architectures have been shown to be effective in multiple molecular tasks including variants of the SchNet and DimeNet architectures [9, 10].

In addition to publishing the dataset used in this study, the Open Catalyst Project implemented successful model architectures for this task and reported a set of scoring metrics. Table 1 shows the model architecture, number of trainable parameters, and the mean overage error (MAE) on the forces task. The values reported are from training on the smallest available training set (200K observations).

| Architecture | Number of Trainable Parameters | Forces MAE ↓ |
|---|---|---|
| *SchNet* | 6,000,000 | 0.0814 |
| *DimeNet++* | 12,000,000 | 0.0741 |

Table 1: Size and performance of molecular GNN models on this task reported in the initial Open Catalyst Project paper [1].

Since this initial analysis suggested that performance was increasing with larger models, recent work has focused on scaling these models to over 1 billion parameters and training for over 11,000 GPU hours to increase performance on this task [4, 5]. This effort has largely been focused on variants of the GemNet architecture which incorporates information about the 3-D Structure of the atomic system by embedding the angles between triplets and quadruplets of atoms [11]. Notably, the GemNet-dT variant has achieved a forces MAE of 0.0693 on this task but contains over 31 million parameters.

However, there is a gap in the current literature surrounding this problem related to the development of more lightweight approaches. Making use of smaller models can make this problem more approachable for machine learning practitioners around the world. This would allow a larger, more diverse set of researchers to work on this problem. Lightweight methods would also allow for more rapid testing of the efficacy of different design choices that could later be scaled up for larger models.

## 3. Methods

In this section, several methods are discussed for framing this problem in a way that can be solved by deep learning including how graphs are constructed from the data. Lightweight GNN modeling approaches and architecture design patterns will also be discussed.

### 3.1. Data Processing

Given that molecules are comprised of multiple atoms held together by chemical bonds, it is natural to adopt graph theoretical approaches to representing them where nodes represent the atoms and edges represent bonds.

In the context of this problem, the aim is to represent the entire adsorbate-surface system as a graph. This includes interactions between atoms that are nearby, but not necessarily bonded. To generate the most meaningful graph representation of the system, we construct a radius graph where edges represent nearby interaction between pairs of atoms [1]. Figure 2 illustrates the graph construction process. The structure is repeated in multiple tiles (e.g., (-1, 0, 0) and (0, 0, 0)) to allow the full 3-dimensional system to be represented. As a result, unique relative distance and edge features can be encoded in the graph representation of the system [6]. This graph representation encodes 3-dimensional structural information in a way that likely could not be captured in a tabular representation.

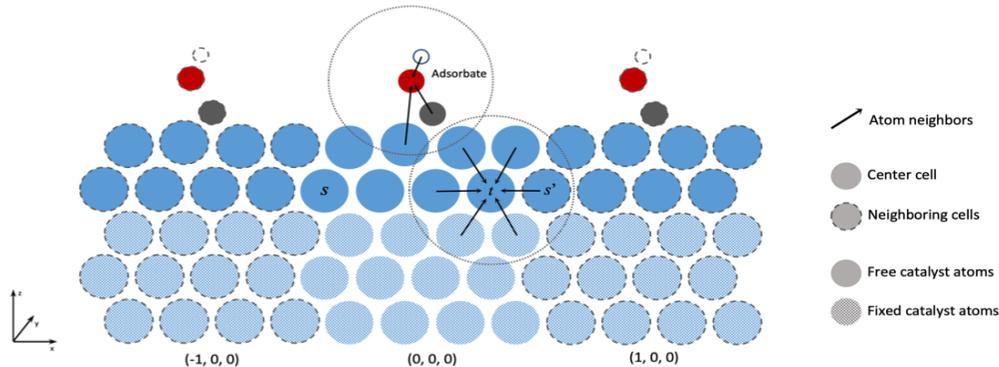

Figure 2: Constructing graph representation of adsorbate-surface system from [6]. Directed edges are drawn between atoms within cutoff distance from one another, so edges are always bidirectional. Since the nodes are periodically repeated, two atoms may have multiple directed edges if they lie within the cutoff distance in repeated cells, resulting in unique relative distance and edge features.

## 3.2. Graph Neural Networks

With each observation in the dataset represented as a graph describing the 3-dimensional structure of the adsorbate-surface system, this problem is well-suited for modeling with graph neural networks. These models are typically constructed using a combination of design patterns that are proven to work well in similar domains. In this study we explore several successful design choices of large GNN-based molecular models and look to scale them to a fraction of the size while achieving comparable accuracy. Some of the design choices explored include gaussian expansion, message passing, geometric message passing, and symmetric message passing.

**Gaussian Expansion:** Gaussian expansion is a method for representing a graph in a continuous, non-sparse space rather than a discrete set of nodes and edges [12]. This has been found to more amenable to learning meaningful representations in GNNs, particularly in molecular applications [9, 11, 12].

**Message Passing:** Message Passing GNNs learn by exchanging information between adjacent nodes in a graph. This involves each node sending and receiving information from its neighbors and using that information to update the node's internal representation of the graph. This process allows the model to learn the complex structure and relationships within a graph which can be used to complete the desired task.

**Geometric Message Passing:** In addition to traditional message passing GNNs, a more robust message passing framework can be implemented to help the model learn a better representation of the 3-dimensional structure which is important for this task. In practice, we can do this by embedding the angles between triplets and quadruplets of atoms in the adsorbate-surface system. Figure 3 illustrates the added angular embeddings for geometric message passing [11].

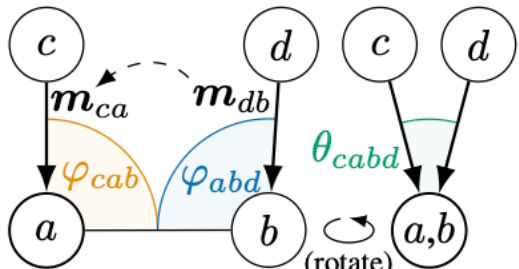

Figure 3: Angles embedding in geometric message passing to help capture the 3-dimensional geometry of the adsorbate-surface system [11].

**Symmetric Message Passing:** Since the graph representation of the adsorbate-surface system consists only of bidirectional edges as described in figure 2, we can adopt symmetric message passing to limit the number of updates required during training. This method has also been shown empirically to yield significant improvements in accuracy for this task [11].

## 3.3. Scaling Down Large Models

Variants of the GemNet architecture, make use of many of the design choices highlighted in the previous section. However, current implementations of these models contain up to 1 billion trainable parameters [5]. This means these models require substantial computational resources and take a long time to train.

However, like large deep learning models from other domains like VGG16 and ResNet-18 from the computer vision literature, the GemNet models contain repeated blocks consisting of the same layers. We can scale down these models by decreasing the depth (number of blocks) and decreasing the embedding size of nodes and edges. Additionally, implementing symmetric message passing as described in Section 3.2 will decrease the size of the model as the edge embeddings are shared between two atoms rather than having separate embeddings for edges in both directions.

In this study, given our computational resources, we will look to train models with less than 5 million parameters while achieving similar accuracy as models that are orders of magnitude larger.

## 4. Experiments

In this section, the design choices highlighted in Section 3.2 were implemented to train several lightweight GNN models and compare accuracy with the current large, state-of-the-art models.

Table 2 shows the features associated with each node and node-pair contained in the dataset. These will be used across all models discussed in this section to predict the per-atom forces experienced during the adsorbate-surface interaction.

| Inputs | |
|---|---|
| **Per atom** | |
| atom positions | 3D $(x, y, z)$ positions |
| group number | column in the periodic table (1-18) |
| period number | row in the periodic table (1-9) |
| electronegativity | tendency to attract electrons (0.5-4.0) |
| covalent radius [113] | size of an atom part of one covalent bond (25-250 pm) |
| valence electrons | # of outer shell electrons (1-12) |
| first ionization energy | energy needed to remove outermost electron (1.3-3.3 eV) |
| electron affinity | energy released when electron attaches to a neutral atom to form a negative ion (−3-3.7 eV) |
| block | s, p, d, f |
| atomic volume | 1.5-4.3 cm$^3$/mol |
| **Per pair** | |
| bond type | single, double, etc. (one-hot) |
| distance | 3D $(x, y, z)$ delta positions |

Table 2: Features for each node and node-pair contained in the dataset without features geometric message passing.

### 4.1. Simple Message Passing GNN with Gaussian Expansion (MPGNN-Tiny)

To start, a simple message passing graph neural network was implemented to establish a baseline for performance without using design characteristics like geometric message passing or symmetric message passing.

This model contained a node embedding layer, an edge embedding layer with gaussian expansion. Both embedding layers used an embedding dimension of 256. The edge features were summed for each node and then passed through two fully connected layers before projecting down to the 3-element output force vector for each node (x, y, z). The sigmoid linear unit (SiLU) activation function was used to add nonlinearity. The model was trained with a batch size of 8 for 50 epochs using the AdamW optimizer. Learning rate scheduling and early stopping were used for regularization. This simple model had 185,000 parameters.

### 4.2. Incorporating Geometric Message Passing (GemNet-Mini)

In addition to the simple model outlined in Section 4.1, we also trained a model that is a bit more complicated, taking advantages of design characteristics like geometric message passing and symmetric message passing.

This model contained a node embedding layer, an edge embedding layer with gaussian expansion, as well as an embedding layer for angle embedding for geometric message passing. These embedding layers had dimension of 256, 256, and 64, respectively. The embeddings are passed through an interaction block before going through different message passing blocks depending on whether it is an atom self-interaction, one-hop geometric interaction, or two-hop geometric interaction. Figure 4 illustrates this architecture as a single block from the GemNet architecture [11].

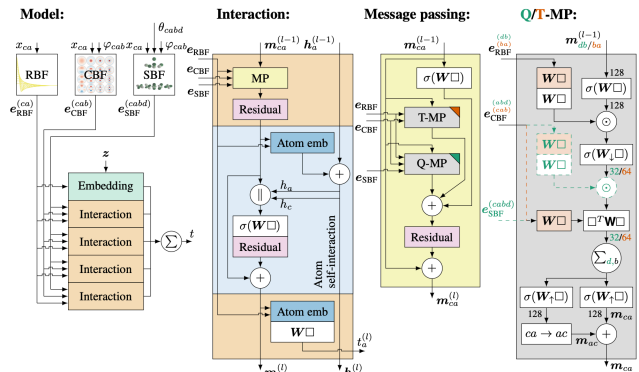

Figure 4: Single GemNet block. ∥ denotes concatenation, and σ a non-linearity. Directional embeddings are updated using three forms of interaction: Two-hop geometric message passing (Q-MP), one-hop geometric message passing (T-MP), and atom self-interactions.

The SiLU activation function was added for nonlinearity and the model was trained with a batch size of 16 for 50 epochs using the AdamW optimizer. Once again, learning rate scheduling and early stopping were used for regularization. The final model had 3.3 million parameters: 1/10 of the size of the smallest published GemNet model.

## 4.3. Results

Table 3 summarizes the results of training these lightweight GNN architectures in terms of the number of trainable parameters and the mean average error (MAE) for predicting the per-atom forces on a validation set of 40K observations. Current state-of-the-art models on the smallest training split for this task (200K observations) are shown for reference. Some larger models like GemNet-XL (1 billion parameters) were not included as they have not reported accuracy on this training split.

| Architecture | Number of Trainable Parameters ↓ | Validation Forces MAE ↓ |
|---|---|---|
| GemNet-dT | 31,000,000 | 0.0693 |
| DimeNet++ | 12,000,000 | 0.0741 |
| SchNet | 6,000,000 | 0.0814 |
| **GemNet-Mini*** | **3,300,000** | **0.0748** |
| **MPGNN-Tiny*** | **185,000** | **0.0827** |

Table 3: Results of experiments relative to current state-of-the-art models. Median baseline Forces MAE is 0.0942. *Denotes models trained in this study.*

Looking at the results summarized above, we can see that these approaches were successful in training models that can reach comparable accuracy values with a fraction of the trainable parameters, allowing them to be trained much faster and using less computational resources.

Notably, our GemNet-Mini architecture (outlined in Section 4.2) achieved a forces MAE of .0748, a 21% relative improvement over median baseline. We can see that this is comparable to the DimeNet++ architecture containing nearly 4X as many parameters as our GemNet-Mini model and a significant improvement over the SchNet model containing nearly 2X as many parameters.

Additionally, we can see that our MPGNN-Tiny model with only 185,000 parameters was able to reach a forces MAE of .0827, an 11% relative improvement over median baseline. We can see that this comparable to the performance of the SchNet architecture containing over 30X as many parameters.

While the larger GemNet-dT architecture outperforms our smaller method, these findings suggest that larger models do not always perform better for this task. Additionally, the success of our GemNet-Mini model relative to other model architectures like SchNet and DimeNet++ suggest that training smaller models can be used to test design patterns that could later be scaled to larger models by teams with more resources.

## 5. Contributions to the Literature

The results of this study add several potential contributions to the literature.

First, we produced a model, GemNet-Mini, that reaches comparable performance to some of the state-of-the-art models for this task with a fraction of the parameters. This could provide smaller teams of researchers with a model that they can use out-of-the-box or fine-tune for a similar task, without significant computational cost unlike the other models for this task. Additionally, our study suggests that smaller models can perform well in this task. This could encourage researchers from a wider set of backgrounds and institutions to work on this problem as it is no longer seen as one that is simply dominated by the largest number of parameters or GPUs.

Additionally, given the success of GemNet-Mini over SchNet and DimeNet++ and supported by the continued success of the GemNet-dT architecture, it seems that more robust design choices like geometric message passing, provide advantages that scale well with larger models. This suggests that lightweight models can be used to test creative design patterns that could later be scaled up to produce models with the highest accuracy. This once again leaves room for individuals and university researchers to contribute, while large AI research laboratories have dominated the literature to date.

The improvements of our GemNet-Mini model which incorporates geometric and symmetric message passing, over the larger SchNet architecture without these features, suggests that encoding the 3-dimensional geometry of the overall system using angular embeddings is important for good performance on this task. This could provide insights to researchers in the chemistry field to develop theories for understanding some properties about catalysts based on structure. Current theories are rooted in quantum mechanics, hence the complexity of DFT calculations.

## 6. Conclusion

In this work, we investigated methods for understanding the catalytic activity of a material by modeling the per-atom interactions during different adsorbate-surface interactions. Our approach ultimately involved using successful design patterns from graph neural networks across molecular modeling problems and scaling them down to make them more computationally tractable for machine learning practitioners around the world.

We saw that by implementing robust design patterns, like geometric and symmetric message passing, we were able to train models to achieve better performance than some popular model architectures with a fraction of the number of parameters and computational burden. Our GemNet-Mini model architecture was able to reach a forces MAE of 0.0748, similar performance to the DimeNet++ architecture 4X its size. Additionally, our MPGNN-Tiny model reached a forces MAE of 0.0827, comparable to the SchNet model, a popular architecture with over 30X as many parameters. These findings show that more parameters do not always lead to better performance for this task, and that we can make use of smaller models for testing design patterns that can later be scaled to larger implementations. These findings could encourage people from more diverse domains to work on this problem, ultimately leading to faster progress.

This study also poses avenues for future studies. In addition to training smaller model architectures from scratch, recent developments have been made in compression of large pre-trained models. Methods for model compression include pruning, quantization, and knowledge distillation. These methods could be used to take large pre-trained models with high observed performance and create smaller models [13]. This could be particularly helpful for researchers who would like to use these models in place of DFT, or fine-tune them for other molecular tasks, but don't possess adequate computational resources to handle a model with 1 billion parameters.